\documentstyle[12pt]{article}

\begin{document}

\topmargin 0pt
\oddsidemargin 5mm

\setcounter{page}{1}
\vspace{2cm}
\begin{center}
{\bf Spherical P-spin glass at $P\to\infty$ and the information storing
 by continuous spins}\\
\vspace{5mm}
{\large  D.B. Saakian }\\
\vspace{5mm}
{\em Yerevan Physics Institute}\\
{Alikhanian Brothers St.2, Yerevan 375036, Armenia\\
and
Laboratory of Computing techniques and automation, JINR,
141980 Dubna, Russia\\
Saakian @ vx1.YERPHI.AM}
\end{center}

\vspace{5mm}
\centerline{\bf{Abstract}}

The $P\to\infty$ limit was considered in the spherical P-spin glass.
It is possible to store information in the vacuum configuration of
ferromagnetic phase. Maximal allowed level of noise was calculated in
ferromagnetic phase.\\

\vspace{5mm}

Derrida's model [1,2] has been applied for optimal coding [3-6]. One chooses 
couplings $J_{i_1..i_p}^0$ for the N spin hamiltonian  $H(\sigma )$ to have 
single and given vacuum configuration $\{\xi_i\}$:  \begin{equation} 
J_{i_1..i_p}^0=\xi _{i_1} \dots \xi _{i_P} \end{equation} \begin{equation} 
H=-\sum_{1\leq i_1<i_2..<i_p\leq N} \sum _{k=1}^{\alpha N} C^k_{i_1..i_p} 
J_{i_1..i_p}^0\sigma_{i_1}..\sigma_{i_p}
\end{equation}
where $C^k_{i_1..i_p}$ is a connectivity matrix. It has only one nonzero
element (equal to 1) at any $k$ for some  choice of indices $ (i_1 \cdots i_P)$.
Spins $\sigma _i, \xi _i$ taking values $\pm1$. Our hamiltonian $H(\sigma )$
has a minimum at the configuration $\{\sigma _i\}=\{\xi _i\}$.\\
Our original message (with length N) 
As has been proven in [4], it stays  a true vacuum, even when one makes our
couplings noisy ( our original couplings $J_{i_1..i_p}^0$ with the
probability  $\frac{1+m}{2}$ stay correct and with probability
$\frac{1-m}{2}$ change their sign), if
\begin{equation}
\alpha [\ln 2+\frac{1+m}{2}\ln \frac{1+m}{2}+\frac{1-m}{2}\ln \frac{1-m}{2}]
\ge \ln 2
\end{equation}
This inequality, derived as a condition for the existence of ferromagnetic
phase of the model (1),(2) in the limit $P\to\infty$ coincides with Shannon
inequality from information theory. Only Derrida's hamiltonian saturates
Shannon limit. For the other choice of hamiltonian $H(\sigma)$ one needs in
more couplings (than $\alpha N$ from (3)) to support ferromagnetic phase
with the full magnetization $<\sigma_i\xi_i>\sim 1$.\\
Our original message $\{\xi_i\}$ (with length N) was transformed into encoded 
message (with length $\alpha N$) $\{ J^0_{i_1..i_P}\}$. So we performed coding.
One can extract(decoding) our original message $\{\xi_i\}$ (decoding) by vacuum search of 
hamiltonian (even with noisy couplings). When with probability close to $1$ 
vacuum of this hamiltonian is our old one, our coding scheme is succesfull.
On the language of statistical physics we need in a full magnetization (at low
temperatures) at our given configuration. To suppress finite size corrections
we need in large $P$.\\ 
Is it possible to construct similar hamiltonian for the continuous spins?\\
Let us consider $N$ spins $\sigma _i$ under constraint of
\begin{equation}
\sum_{i=1}^N(\sigma_i)^2=N
\end{equation}
and the same for the $\xi _i$. Again we consider hamiltonian (1),(2).
If our connectivity matrix has been chosen
symmetric by $i _\alpha$ and one enlareges sum by indices $i_\alpha$ in (2)
till $1 \leq i _\alpha \leq N$ , then $H(\sigma )$ goes to
\begin{equation}
H(\sigma )=-(\sum_i \sigma_i \xi_i)^P\frac{\alpha N}{N^P}
\end{equation}
This function has a minimum at $\{\sigma _i\}=\{\xi _i\}$. In the limit
 $N\gg P\gg 1$ it will be the minimum of the hamiltonian (2) with the
constraint sum for indices in (2) (instead of $1 \leq i _\alpha \leq N$).\\
To search something similar to inequality (3) one needs to find phase
structure of spherical P-spin glass [7-10]. The static limit of model was
solved in [8], we need in a little modification of their calculation to
consider our case $P\to\infty$ and ferromagnetic phase.\\
We see, that our hamiltonian has a correct minimum, even in the class of
continuous spins with spherical constraint for the homogeneous choice of
connectivity matrix $C$. The situation is different from the case of discrete
spins, where the condition of homogeneity is not necessary.\\
The situation will be the same for NN models (again one could look for minimum
in the class of continuous spins with spherical constraint).\\
To solve the case of nontrival connectivity matrix we need in a solution of
 diluted spherical P-spin glass.
As a first step let us consider the simple case of fully connected model 
(matrix C disappears).\\
Let us consider hamiltonian
\begin{equation}
H=-\sum_{1\leq i_1<i_2..<i_p\leq N}
(J_0 N/C_N^P+J_{i_1..i_P}\sqrt{N/C_N^P})\sigma _{i_1}..\sigma_{i_p}
\end{equation}
where $J_0$ -ferromagnetic coupling and quenched couplings
$J _{i_1 \cdots i_P}$ (noise) have a $0$ mean and  variance
$<(j_{i_1 \dots i_P})^2>=J^2/2$. The signal/noise ratio  $ J_0/J$ is similar
to $m$ in (4).\\
Calculating $Z ^n$ by means of replica trick , as in [8], we derive
\begin{eqnarray}
Z^{n}=\int_{-i\infty}^{i\infty}
\prod_{\alpha<\beta}
\frac{NdQ_{\alpha\beta}d\lambda_{\alpha\beta}}
{2\pi}\prod_{\alpha}\int_{-i\infty}^{i\infty}\frac {\sqrt N}{2\pi }
d\lambda_{\alpha \alpha}\int_{-i\infty}^{i\infty}\frac{dt_{\alpha}dm_{\alpha}}
{2\pi }\exp (NG)
\end{eqnarray}
where
\begin{eqnarray}
G=J_0B\sum_{\alpha}(m_{\alpha})^P+\frac{B^2J^2}{4}
\sum_{\alpha,\beta}(q_{\alpha,\beta})^P
-\frac{1}{2}\sum_{\alpha,\beta}q_{\alpha,\beta}\lambda_{\alpha,\beta}
\sum_{\alpha}t_{\alpha}m_{\alpha}+\frac{1}{2}\ln 2\pi+\nonumber\\
\ln\int_{-\infty}^{\infty}\prod_{\alpha}dx_{\alpha}\exp \{\frac{1}{2}
\sum_{\alpha,\beta}\lambda_{\alpha,\beta}x_{\alpha}x_{\beta}+
\sum_{\alpha}t_{\alpha}x_{\alpha}\}
\end{eqnarray}
In this expression $m_\alpha=<x_\alpha>, q_{\alpha \beta}=<x_\alpha x_\beta>$
, $t_\alpha$ and $\lambda_{\alpha \beta}$ are conjugate for them. Equation
(4) gives $q_{\alpha \alpha}=1$.\\
After integrating by $x_\alpha,  t_\alpha$, we have
\begin{eqnarray}
G=\frac{1}{2}\ln 2\pi+J_0B(m)^P+\frac{B^2J^2}{4}
\sum_{\alpha,\beta}(q_{\alpha,\beta})^P+
\frac{1}{2}\sum_{\alpha,\beta}\lambda_{\alpha,\beta}
(m_\alpha m_\beta-q_{\alpha,\beta})-\nonumber\\
\frac{1}{2}\ln det\{-\lambda^{-1}\}_{\alpha,\beta}
\end{eqnarray}
The saddle point condition for the $\lambda_{\alpha, \beta}$ gives
$q_{\alpha,\beta}=m_\alpha m_\beta-\{\lambda^{-1}\}_{\alpha,\beta}$\\
At the high temperetures the system lives in the paramagnetic phase, where
$m_\alpha=0,q_{\alpha\ne \beta}=0$ . It is easy to derive, as in [8]
\begin{equation}
G=\frac{1}{2}(1+\ln 2\pi)+\frac{B^2J^2}{4}
\end{equation}
In the SG phase we need in one-level breaking of replica symmetry, to
block of indices of order $m_1$. We have $q_{\alpha \alpha}=1,
q_{\alpha \beta}=q$ for $\alpha \beta$ from one block of order $m_1$,
and for other indice sets $q_{\alpha \beta}=0$.\\
Then (9) goes to
\begin{equation}
G=\frac{1}{2}(1+\ln 2\pi)+\frac{B^2J^2}{4}[1+(m_1-1)q^P]+\frac{m_1-1}{2m_1}
\ln [1+(m_1-1)q]
\end{equation}
Taking derivatives by $m_1$ and  $q$ and dividing equatios we derive system
\begin{equation}
\label{L16}
\left\{
\begin{array}{l}
\frac{q^2}{p}=\frac{(1-q)(1-q+mq)}{m^2}\ln{ (1-q+mq)}{(1-q)}-
\frac{q(1-q)}{m}\\
\frac{PB^2J^2}{4}q^P=\frac{1}{2(1-q)(1-q+mq)}
\end{array}
\right.
\end{equation}
Let us consider limit $P\to\infty$. It is a reasonable anzats
\begin{equation}
q=1-\epsilon ,\epsilon\to 0
\end{equation}
Then (16) goes to
\begin{equation}
\label{L21}
\left\{
\begin{array}{l}
\frac{1}{P}=
\frac{\epsilon}{m}\ln \frac{m}{\epsilon}\\
\frac{PB^2J^2}{2}=\frac{1}{\epsilon m}
\end{array}
\right.
\end{equation}
Its solution is
\begin{eqnarray}
\epsilon=\sqrt{\frac{2}{\ln P}}\frac{1}{PBJ}\\
m=\frac{\sqrt{2\ln P}}{BJ}
\end{eqnarray}
For the free energy we have
\begin{equation}
\label{L24}
G=\frac{BJ\sqrt{2\ln P}}{2}-\frac{1}{2}\ln \frac{PBJ}{\sqrt{2\ln P}}
+\frac{1}{2}(1+\ln 2\pi)
\end{equation}
We find $B_c$ for the phase transition (paramagnetic-SG) by comparing (24)
and (13):
\begin{equation}
\label{L25}
B_c=\sqrt{2\ln P}/J\\
\end{equation}
For the ferromagnetic phase we have in the bulk approximation (just enough
for our purposes) $\ln Z/N=J_0B$. More accurate consideration gives
\begin{equation}
\label{L44}
\frac{\ln Z}{N}=J_0B-\frac{\ln P}{2}
\end{equation}
Comparision with two other phases gives, that at high $B$
ferromagnetic phase appears, when
\begin{equation}
\left\{
\begin{array}{l}
J_0>\sqrt{\frac{\ln P}{2}}J\\
B>B_c\equiv\sqrt{2\ln P}/J
\end{array}
\right.
\end{equation}
It is the main result of this work. It resembles the result for the
discrete spins with Potts like interaction and color equal to $P$.
It will be very interesting to solve the
case of diluted couplings, as well as to consider similar models for the
pattern recognition like Hopiefild or Gardner. The distance in the space
of pattern with our spins will be like euclidean and as a consequence,
closely connected with the reality, than artifical discretization by means of
$Q$ color spins.\\
The multicriticity point has coordinates
\begin{equation}
\left\{
\begin{array}{l}
B_c=\sqrt{2\ln P}/J\\
J_0=\sqrt{\frac{\ln P}{2}}J
\end{array}
\right.
\end{equation}
It will be interesting to search analogy with the Nishimori line. Direct 
enlargment for
the methods of [11] is impossible, but the idea of resonance between Gibbs
partition and inhomogeneous partition of couplings is too beatifull to throw it.
\\{\it Acknowledgments}
I would like to thank Yu Lu for the invitation to ICTP, Ye. P. Jidkov-to JINR.
I am  grateful to  S. Franz for the help, M. Virasoro for a discussion
of SG and H. Nishimori for useful remarks.
This work was supported by  German ministry of Science and Technology
Grant 211 - 5231.

\end{document}